\documentclass[fleqn,usenatbib]{mnras}

\usepackage{amssymb,amsmath,framed,xcolor}

\usepackage{bm}
\expandafter\ifx\csname package@font\endcsname\relax\else
 \expandafter\expandafter
 \expandafter\usepackage
 \expandafter\expandafter
 \expandafter{\csname package@font\endcsname}
\fi
\usepackage[T1]{fontenc}

\DeclareRobustCommand{\VAN}[3]{#2}
\let\VANthebibliography\thebibliography
\def\thebibliography{\DeclareRobustCommand{\VAN}[3]{##3}\VANthebibliography}

\usepackage{graphicx}
\usepackage{bm}
\usepackage{color}
\usepackage{newtxtext,newtxmath}

\title[AGN outflows and habitability]{The impact of AGN outflows on the surface habitability of terrestrial planets in the Milky Way}

\author[Ambrifi et al.]{A. Ambrifi$^{1}$, A. Balbi$^{1}$\thanks{E-mail: balbi@roma2.infn.it}, M. Lingam$^{2,3}$\thanks{E-mail: mlingam@fit.edu}, F. Tombesi$^{1,4,5,6, 7}$ and E. Perlman$^{2}$
\\
$^{1}$Department of Physics, Tor Vergata University of Rome, Via della Ricerca Scientifica 1, I-00133 Rome, Italy\\
$^{2}$Department of Aerospace, Physics and Space Sciences, Florida Institute of Technology, Melbourne, FL 32901, USA\\
$^{3}$Department of Physics, The University of Texas at Austin, Austin, TX 78712, USA\\
$^{4}$INAF - Astronomical Observatory of Rome, Via Frascati 33, I-00078 Monte Porzio Catone, Italy\\
$^{5}$Department of Astronomy, University of Maryland, College Park, MD 20742, USA\\
$^{6}$NASA - Goddard Space Flight Center, Code 662, Greenbelt, MD 20771, USA\\
$^{7}$INFN - Roma Tor Vergata, Via della Ricerca Scientifica 1, I-00133 Rome, Italy\\
}
\date{Accepted XXX. Received YYY; in original form ZZZ}

\pubyear{2021}

\begin{document}
\label{firstpage}
\pagerange{\pageref{firstpage}--\pageref{lastpage}}
\maketitle

\begin{abstract}
It is well-known that active galactic nuclei (AGN) are accompanied by winds and outflows, some of which may reach weakly relativistic speeds of about $10$ percent the speed of light. Yet, in spite of their ubiquity, the impact of AGN outflows in modulating surface habitability of terrestrial planets on galactic scales, using the Milky Way as the basis for comparison, is poorly investigated and inadequately understood. In this work, we address this issue by focusing on two key mechanisms: AGN winds can heat atmospheres and drive atmospheric escape, as well as stimulate the formation of nitrogen oxides and thence cause ozone depletion. By developing simple models, we estimate the maximal distance up to which these deleterious effects are rendered significant for Earth-like planets in the Milky Way, and thereby demonstrate that this value may extend to $\lesssim 1$ kpc. In the case of quasars hosting larger supermassive black holes, such effects could actually influence the AGN host galaxy as a whole.
\end{abstract}




\section{Introduction}
With the rapid advent of (exo)planetary science, the interest in gauging and quantifying the surface habitability of terrestrial planets has grown commensurately \citep{SS13,CBB16,MB18,ML21}. A bevy of publications in the twenty-first century have demonstrated that habitability is not only modulated by endogenous planetary processes but also by stellar \citep[e.g.,][]{TBM07,LL19,ABC20} and galactic \citep{Gon05,NP08,GM18} astrophysical mechanisms.

In the past few years, there has been a flurry of activity directed toward ascertaining the impact of the central supermassive black hole (SMBH) on habitability on galactic scales, encompassing both active galactic nuclei (AGN) \citep{BT17,FL18,CFL18,LGB,AC19,WKB,LCD20} as well as tidal disruption events (TDEs) \citep{PBL20}. However, these studies delved exclusively into the positive and negative ramifications of electromagnetic radiation (e.g., X-rays, ultraviolet radiation, and optical light) emitted during the peak activity of the central SMBH.

In actuality, however, the ubiquity of outflows and winds (which are used interchangeably herein) associated with active galactic nuclei is thoroughly documented from an observational standpoint and backed by theoretical modelling \citep{Krol99,Mer13,KP15,HCT18,VMB20}. In particular, one notable class of outflows, the ultra-fast outflows (UFOs), are known to attain weakly relativistic speeds of $\sim 0.1 c$ \citep[e.g.,][]{TCR10,TCR11,GRT13,TTM14,TMV15,CCV21}. 
Similarly powerful outflows have been found in the broad absorption line (BAL) quasars \citep[e.g.,][]{rankine20,xu19,wey91,hfol03}, a significant number of which belong to the UFO class, which also includes the less-powerful Seyfert galaxies. We do not consider jets in this paper, because while they are considerably faster \citep{lister21}, their highly collimated nature means that they affect a much smaller region of the AGN's host galaxy.

Despite the commonality of AGN outflows, surprisingly few publications have attempted to assess their role(s) in regulating habitability. In fact, the studies in this respect date from the 1980s \citep{Cla81,PAL83,LaV87} and were of a semi-quantitative character. Furthermore, the field of AGN outflows has advanced by leaps and bounds since this period owing to a combination of empirical and theoretical breakthroughs \citep{Krol99,KH13,Mer13,HCT18}. In this same period, considerable progress has been accomplished in comprehending the multifarious physical processes (e.g., non-thermal atmospheric escape) that shape the habitability of planets and moons \citep{LBC09,CBB16,ABC20,ML21}. Based on the progress made in AGN phenomenology and physics, the time is arguably ripe to reevaluate the consequences of AGN winds vis-\`a-vis governing habitability, which constitutes the chief rationale for our paper.

Moreover, it is important to recognise that analyses of the effects of central black hole activity on habitability are not merely interesting in their own right but also because they can be self-consistently incorporated into state-of-the-art numerical simulations of galactic habitability \citep{DCR15,GH16,VSM16,FDC17,SGM17,SHL18,DVC19,SVM19,SGH21} -- which have built on earlier works \citep[e.g.,][]{GBW01,LFG04} -- and include high-energy phenomena such as supernovae or gamma-ray bursts. These statistical models enable us to trace how the habitability of the Universe as a whole has evolved with cosmic time.

The general outline and structure of the paper is as follows. For starters, in Section \ref{SecAGNPhen}, we furnish the relevant background material concerning AGN winds required for our treatment. Next, in Section \ref{SecAtmos}, we look at how AGN outflows may contribute to atmospheric heating and escape, and the capacity of AGN winds to cause ozone depletion is elucidated in Section \ref{SecOzone}. Finally, we conclude with an exposition of our results in Section \ref{SecConc}.

\section{AGN outflows: phenomenological details}\label{SecAGNPhen}
In this section, we provide some of the salient details concerning the outflows/winds from AGN that are employed hereafter. Comprehensive summaries of the phenomenological details of AGN can be found in \citet{Krol99}, \citet{AH12}, \citet{Mer13}, and \citet{Net15}, whereas the specifics of AGN outflows are explicated in the reviews by \citet{KP15}, \citet{HCT18}, and \citet{VMB20}.

For a spherically symmetric outflow \citep[e.g.,][]{LRR21}, which is the scenario that we investigate here, the mass outflow rate $\dot{M}_\mathrm{out}$ is expressible as
\begin{equation}\label{dens}
    \dot{M}_\mathrm{out}=4\pi b R^2 \rho v_\mathrm{out},
\end{equation}
where $b$ represents the fraction of solid angle encompassed by the outflow, $R$ is the distance from the centre of the Milky Way, and $\rho$ and $v_\mathrm{out}$ are the density and velocity of the outflow at the given location, respectively. From here onward, we will work with $b \approx 1$, since it represents a reasonable assumption used in several theoretical publications on AGN feedback from quasi-spherical outflows \citep[e.g.,][]{KP03,KP15,ZK12,FQ12}. This value is consistent with the high fraction of AGN with detected ionised winds and with the large opening angle of UFOs estimated in the literature \citep[e.g.,][]{TMV15,nardini2015,fiore2017,LRR21}. Moreover, considering our Milky Way, the multi-wavelength observations of the quasi-spherical ``Fermi Bubbles'' are most likely connected to an enhanced past activity of Sagittarius A* \citep[e.g.,][]{zubo2011,zubo2012}.

The Eddington luminosity associated with the black hole is
\begin{equation}
    L_\mathrm{Edd} = \frac{4\pi G c m_p}{\sigma_T}M_\mathrm{BH} \approx 3.3 \times 10^4\,\left(\frac{M_\mathrm{BH}}{M_\odot}\right) L_\odot
\end{equation}
where $m_p$ signifies the proton mass, $\sigma_T$ is the Thomson scattering cross-section, and $M_\mathrm{BH}$ is the black hole mass. The bolometric luminosity of the AGN ($L_\mathrm{AGN}$) is expressible in terms of the Eddington ratio parameter $\lambda_\mathrm{Edd}$ as follows:
\begin{equation}\label{EddRat}
    \lambda_\mathrm{Edd} \equiv \frac{L_\mathrm{AGN}}{L_\mathrm{Edd}}.
\end{equation}
Based on observations, a substantial fraction of all AGN exhibit the canonical value of $\lambda_\mathrm{Edd} \approx 1$ \citep{MRG04,SE10,BVB17,ACG18}; we shall adopt this fiducial estimate henceforth.

\subsection{Specific approach and relevance to AGN}
As there are a multitude of outflows and winds associated with AGN, we will focus on UFOs (UFOs) -- as they represent the outflows/winds with the highest velocity and power -- and model them by utilising the prescription and approach in \citet{KP03}; see also \citet{KP15} and \citet{HCT18}. We presume that these winds are accelerated due to the electromagnetic radiation emitted during the accretion process, with the latter exerting its radiation pressure on an optically thick medium. The momentum rate of the outflow ($\dot{p}_\mathrm{out}$) is written as
\begin{equation}\label{MomOutF}
    \dot{p}_\mathrm{out} \equiv \dot{M}_\mathrm{out}v_\mathrm{out} \approx \frac{L_\mathrm{AGN}}{c} \approx \frac{\lambda_\mathrm{Edd}L_\mathrm{Edd}}{c},
\end{equation}
where we have assumed that the outflow speed remains approximately constant, and the second equality is obtained by drawing on (\ref{EddRat}). Note that the above equation implies that the momentum rate of the UFO is similar to that of the electromagnetic radiation -- this relationship has been extensively observed for UFOs \citep[e.g.,][]{TCRB,TCRN13,GRM15,FFC15,HTF16,FMC19}.

Next, we turn our attention to the kinetic power associated with the outflow, which we denote by $\dot{E}_k$. From the definition of the kinetic energy, it is apparent that
\begin{equation}\label{KinPow}
    \dot{E}_k \equiv \frac{1}{2}\dot{M}_\mathrm{out}v_\mathrm{out}^2.
\end{equation}
On substituting (\ref{MomOutF}) into (\ref{KinPow}), we end up with
\begin{equation}\label{Ekint}
    \dot{E}_k \approx \frac{v_\mathrm{out}}{2c}L_\mathrm{AGN} \approx 0.05\, \lambda_\mathrm{Edd}L_\mathrm{Edd} \left(\frac{v_\mathrm{out}}{0.1 c}\right)
\end{equation}
where the second equality follows from the fact that $v_\mathrm{out} \approx 0.1 c$ for UFOs \citep{TCR10,TCR11,GRT13,TTM14}. Next, on adopting the prescribed values for $v_\mathrm{out}$ and $\lambda_\mathrm{Edd}$, we arrive at the result
\begin{equation}\label{Ekfin}
    \dot{E}_k\approx 0.05 \,L_\mathrm{Edd},
\end{equation}
which is generally consistent with observations of outflows with kinetic power that can attain a few percent of $L_\mathrm{AGN}$ \citep[e.g.,][]{CK12,TCRB,TCRN13,FFC15,GRM15,TMV15}.

At this stage, we must address the post-shock winds, whose characteristics are identified using the subscript `$ps$'. The UFOs are anticipated to shock the ambient interstellar medium (ISM), and communicate their momentum and (kinetic) energy to the latter \citep{KP15}. Hence, it is evident that two broad and distinct scenarios could be feasible on physical grounds \citep{KP15}. In the energy-driven case (labelled by the subscript `ed'), the kinetic power of the UFO is efficiently transferred to the post-shock wind and the total wind energy is consequently conserved. In contrast, the momentum-driven outcome (denoted by the subscript `md') corresponds to the one wherein the momentum of the UFO is effectively communicated to the post-shock wind; in this instance, a fraction of the total energy of the UFO is radiated away shortly after the shock has transpired. 

\subsection{Energy-driven post-shock winds}\label{SSecMomDr}
Let us first examine the energy-driven post-shock wind. Since we have posited that the kinetic power is the same as the UFO, the kinetic power of the post-shock wind is
\begin{eqnarray}\label{Eked}
   &&  \dot{E}_{k,ed} \equiv \frac{1}{2}\dot{M}_{ps} v_{ps}^2 \approx  \dot{E}_k \approx \frac{1}{2}\dot{M}_\mathrm{out} v_\mathrm{out}^2 \nonumber \\
    && \hspace{0.4in} \approx 0.05\, \lambda_\mathrm{Edd}L_\mathrm{Edd} \left(\frac{v_\mathrm{out}}{0.1 c}\right) \approx 0.05 \,L_\mathrm{Edd},
\end{eqnarray}
where $\dot{M}_{ps}$ and $v_{ps}$ are the mass outflow rate and the speed of the post-shock wind, respectively. The expressions in the second line were obtained by invoking (\ref{Ekint}) and (\ref{Ekfin}). 

One of the interesting features of the energy-driven post-shock wind is that it evinces a higher momentum outflow rate compared to that of the initial UFO, as illustrated below.
\begin{eqnarray}\label{edradpress}
    &&\dot{p}_\mathrm{ed} \equiv \dot{M}_{ps}v_{ps} \approx \frac{2\dot{E}_{k,ed}}{v_{ps}} \approx \dot{p}_\mathrm{out} \left(\frac{ v_\mathrm{out}}{v_{ps}}\right) \nonumber \\
    && \hspace{0.4in} \approx 33\, \dot{p}_\mathrm{out} \left(\frac{v_\mathrm{out}}{0.1 c}\right) \left(\frac{v_{ps}}{1000\,\mathrm{km/s}}\right)^{-1},
\end{eqnarray}
where we have made use of (\ref{MomOutF}) and (\ref{Eked}) to simplify the expression. The quantity $v_\mathrm{out}/v_{ps} \approx 30$ is called the momentum-boost factor, and the characteristic value has been calculated by adopting $v_\mathrm{out} \approx 0.1 c$ from earlier in tandem with specifying $v_{ps} \approx 1000$ km/s \citep{DQM12,FQ12,ZK12,KP15,TMV15}. The estimated momentum-boost factor is in agreement with several observations of galactic-scale outflows, which have revealed momentum-boost factors of up to $\sim 20$-$60$ \citep{ZK12,CMS14,FFC15}.

\subsection{Momentum-driven post-shock winds}\label{SSecEnDr}
Now, let us consider the momentum-driven post-shock wind. By definition, since the conservation of momentum is valid, the momentum rate of the post-shock wind ($\dot{p}_\mathrm{md}$) becomes
\begin{equation}\label{mdradpress}
    \dot{p}_\mathrm{md} \equiv \dot{M}_{ps}v_{ps} \approx \dot{p}_\mathrm{out} \approx \frac{L_\mathrm{AGN}}{c}
\end{equation}
By availing ourselves of this equation, it is possible to derive the kinetic power ($\dot{E}_{k,md}$) in the momentum-driven scenario, which is therefore estimated to be
\begin{eqnarray}\label{Ekmd}
    && \dot{E}_{k,md} \equiv \frac{1}{2}\dot{M}_{ps}v_{ps}^2 \approx \frac{1}{2}\frac{\lambda_\mathrm{Edd}L_\mathrm{Edd}}{c}v_{ps} \nonumber \\
    && \hspace{0.4in} \approx 0.001\lambda_{Edd}L_\mathrm{Edd} \left(\frac{v_{ps}}{1000\,\mathrm{km/s}}\right) \nonumber \\
     && \hspace{0.4in} \approx 0.001\,L_\mathrm{Edd},
\end{eqnarray}
where the second equality on the right-hand side is derived from (\ref{mdradpress}) and (\ref{EddRat}). The equality in the second line is calculated by inputting the typical velocity of post-shock outflows, to wit, we substitute $v_{ps} \approx 1000$ km/s from the preceding paragraph. The expression $\dot{E}_{k,md} \approx 0.001\,L_\mathrm{Edd}$ is compatible with observations of some cold molecular outflows \citep{FMC19,VMB20}.

In closing, we note that the characteristic lifetime of the AGN is often modelled by the Salpeter timescale \citep{Krol99}, denoted by $\Delta t_\mathrm{Salp}$, which is defined as
\begin{equation}\label{SalpTime}
    \Delta t_\mathrm{Salp} \equiv \frac{M_\mathrm{BH}}{\dot{M}_\mathrm{BH}} \approx \frac{\eta c^2 M_\mathrm{BH}}{L_\mathrm{Edd}},
\end{equation}
where $\dot{M}_\mathrm{BH}$ is the accretion rate and $\eta$ is the radiative efficiency, which has a typical value of $0.1$ \citep{Shen13}. For the canonical choice of $\eta \approx 0.1$, it is known that the Salpeter timescale is around $45$ Myr \citep{Shen13}, implying that we will hereafter deal with $\Delta t_\mathrm{Salp} \sim 10$-$100$ Myr in most instances.

By solving this equation for $\Delta t_\mathrm{Salp} \cdot L_\mathrm{Edd}$, we arrive at
\begin{equation}\label{TotalEn}
  \Delta t_\mathrm{Salp} L_\mathrm{Edd} \approx \eta c^2 M_\mathrm{BH} \approx 7.5 \times 10^{59}\,\mathrm{erg}\,\left(\frac{\eta}{0.1}\right)\left(\frac{M_\mathrm{BH}}{M_\mathrm{SgrA*}}\right),
\end{equation}
where $M_\mathrm{SgrA} \approx 4.1 \times 10^6 M_\odot$ denotes the mass of Sagittarius A* \citep[Table 1]{GCA19}. By construction, note that the left-hand side of this expression is approximately equivalent to the total amount of electromagnetic energy emitted by the supermassive black hole during its AGN phase.  While this parametrisation does not take into account the effects of AGN variability on the outflow luminosity, for the purposes of this paper a rough average is sufficient. Detailed discussions of the temporal variability and evolution of AGN luminosities and their outflows are furnished in \citet{KP15}, \cite{IshFab15}, \cite{Zub18}, and \citet{VMB20}.

\section{Atmospheric heating and escape}\label{SecAtmos}
In this section, we explore the ramifications of UFOs and their post-shock winds in driving atmospheric escape, and briefly elucidate the attendant consequences.

\begin{figure}
\includegraphics[width=8.5cm]{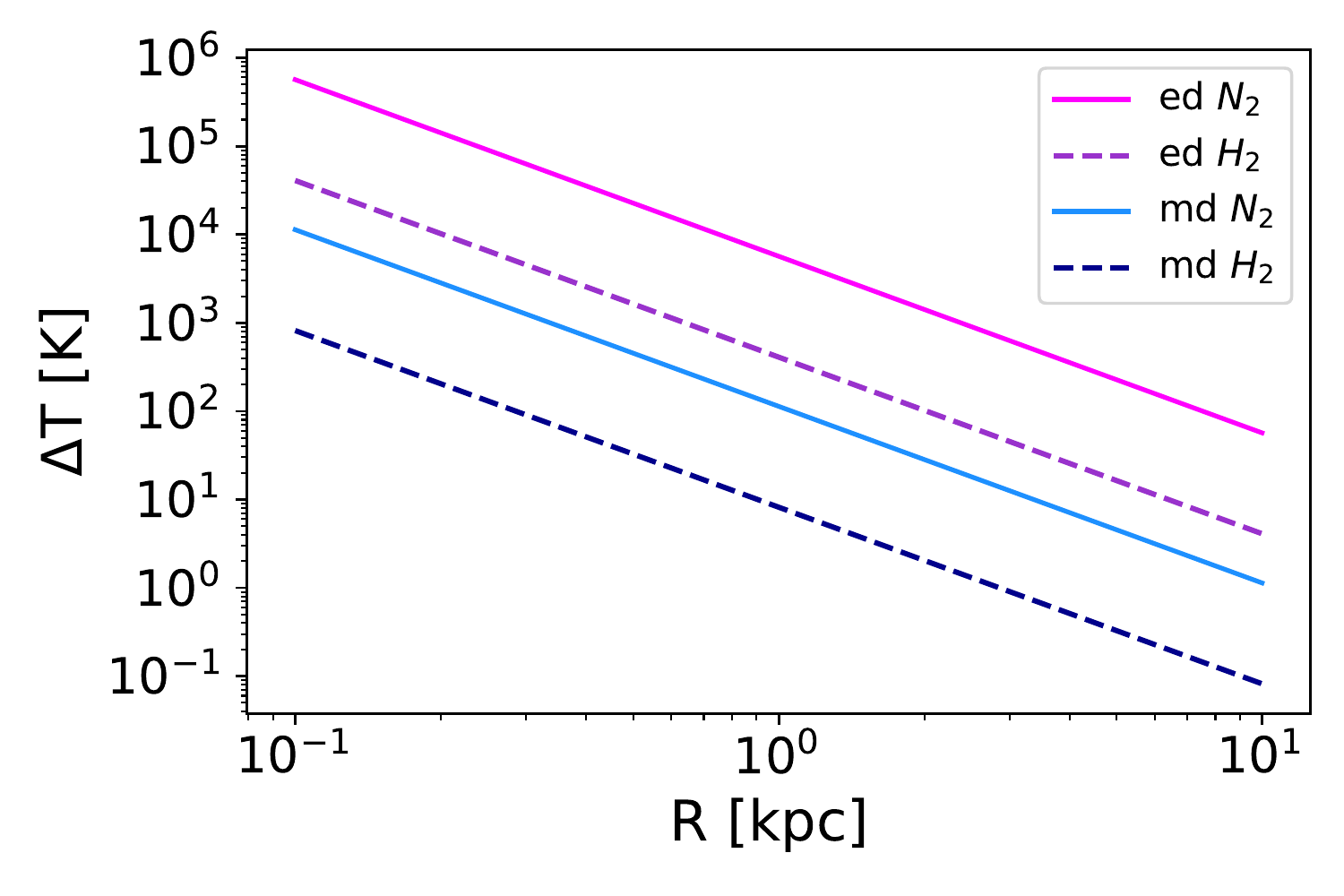} \\
\caption{The maximal increase in atmospheric temperature ($\Delta T$) caused by the energy deposited by the AGN wind as a function of the distance $R$ from the Galactic centre (in kpc). The labels `ed' and `md' refer to the energy- and momentum-driven scenarios, respectively (see Section \ref{SecAGNPhen}). The labels N$_2$ and H$_2$ indicate that the atmosphere is predominantly composed of molecular nitrogen and hydrogen.}
\label{FigHeatAtm}
\end{figure}

\subsection{Temperature-mediated atmospheric escape}
The energy inherent in the AGN outflows is capable of inducing many effects, of which one of the most obvious entails heating of planetary atmospheres. In order to gauge the maximal heating caused and its impact on atmospheric escape, we make the idealised assumption that all the incident energy is used to heat the atmosphere and thus carry out the analysis.

The upper bound on the rise in atmospheric temperature $\Delta T$ driven by the wind is estimated using
\begin{equation}\label{DelTdef}
    \Delta T \approx \frac{Q}{m_\mathrm{atm} \mathcal{C}},
\end{equation}
where $Q$ in the amount of incoming energy, $m_\mathrm{atm}$ is the mass of the atmosphere, and $\mathcal{C}$ represents the specific heat capacity. For Earth, the atmospheric mass is $m_\mathrm{atm} \approx 5.1\times 10^{21}$ g. The total incident energy $Q$ can thence be ascertained by multiplying the incident power with the Salpeter time, which leads us to
\begin{equation}\label{heat}
    Q=\frac{\dot{\mathcal{E}}_k \Delta t_\mathrm{Salp}}{4}\bigg(\frac{R_p}{R}\bigg)^2,
\end{equation}
where $\dot{\mathcal{E}}_k$ is the total kinetic power associated with the outflow, and $R_p$ is the radius of the planet. The factor of $4$ in the denominator on the right-hand side arises because the power distributed over the spherical area of $4\pi R^2$ is intercepted by the planet that is distinguished by a cross-sectional area of $\pi R_p^2$.

Here, it must be recalled that the kinetic power introduced above depends on the specific case under consideration, as elucidated in Section \ref{SecAGNPhen}. When the wind is energy-driven, we have $\dot{\mathcal{E}}_k = \dot{E}_{k,ed} \approx 0.05 L_\mathrm{Edd}$ from (\ref{Eked}), while we specify $\dot{\mathcal{E}}_k = \dot{E}_{k,md} \approx 0.001 L_\mathrm{Edd}$ from (\ref{Ekmd}) for the momentum-driven scenario. In what follows, we will work with $\dot{\mathcal{E}}_k = \epsilon L_\mathrm{Edd}$, with $\epsilon$ embodying the fraction of AGN luminosity converted into the kinetic energy of the wind. For post-shock energy-driven winds and momentum-driven winds we have $\epsilon \approx 0.05$ and $\epsilon \approx 0.001$, respectively.

On substituting (\ref{heat}) into (\ref{DelTdef}), we end up with
\begin{equation}\label{modelheating}
    \Delta T=\frac{\epsilon L_\mathrm{Edd} \Delta t_\mathrm{Salp}}{4 m_\mathrm{atm} \mathcal{C}}\bigg(\frac{R_p}{R}\bigg)^2.
\end{equation}
We evaluate our results for an atmospheric mass equal to that of the Earth's atmosphere. However, we consider two distinct types of atmospheric composition: the first is dominated by N$_2$ and the second by H$_2$. The former is reminiscent of modern Earth, whereas the latter resembles certain super-Earth atmospheres -- which are expected to be potentially common and habitable \citep{ETS08,SBH13,SPP20,MPC21} -- but probably with much lower atmospheric mass. These two examples are merely representative, and we shall not tackle worlds with atmospheric compositions analogous to, say, Venus and Titan; such analyses can comprise the basis of future work.

\begin{figure}
\includegraphics[width=8.5cm]{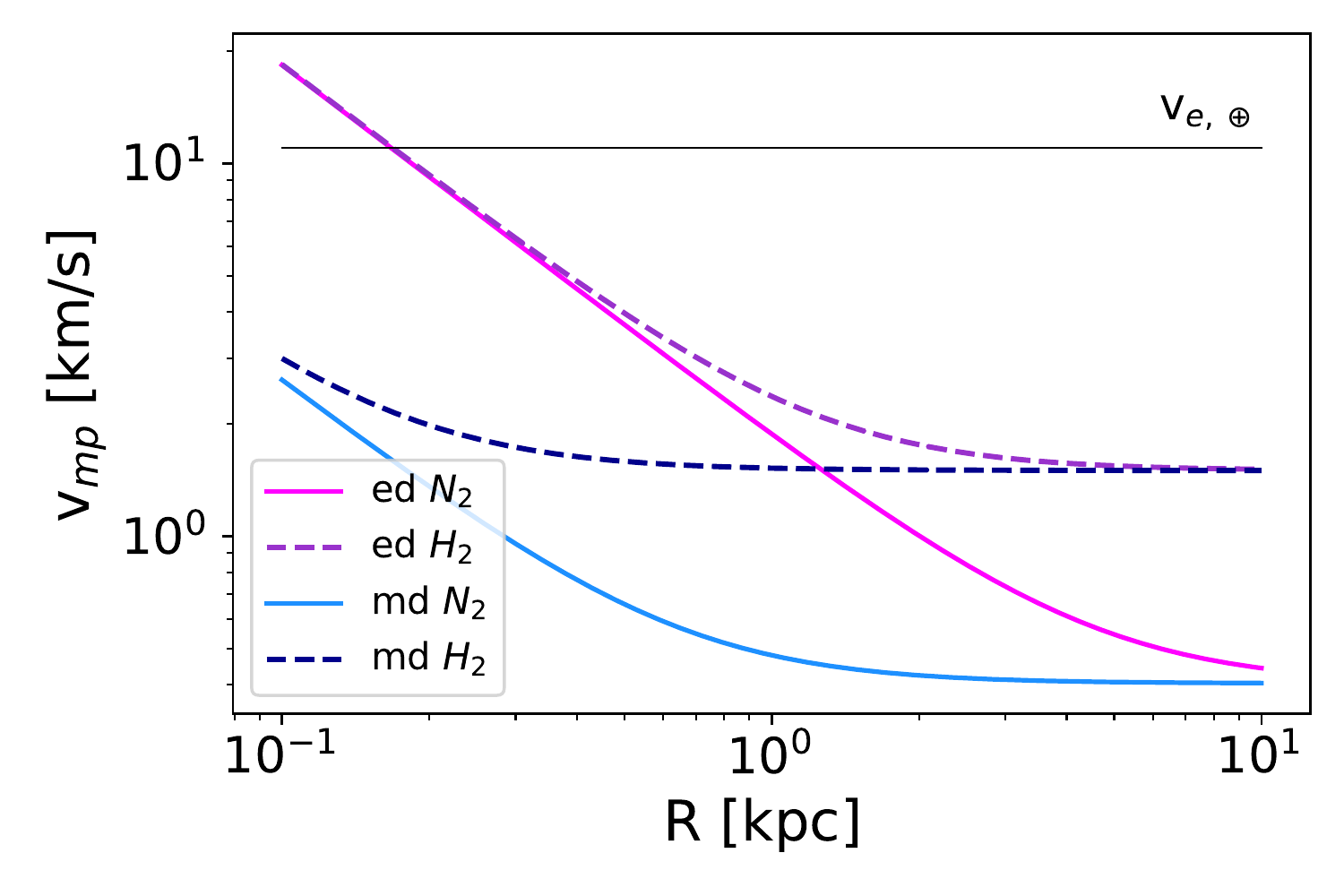} \\
\caption{The most probable speed of molecules in the atmosphere ($v_\mathrm{mp}$) as a function of the distance $R$ from the Galactic centre (in kpc). The The labels `ed' and `md' refer to the energy- and momentum-driven scenarios, respectively (see Section \ref{SecAGNPhen}). The labels N$_2$ and H$_2$ indicate that the atmosphere is predominantly composed of molecular nitrogen and hydrogen. The horizontal line denotes the escape velocity of the Earth.}
\label{FigMPSpeed}
\end{figure}

Using the data from the tables furnished in the NIST Standard Reference Database $^[$\footnote{NIST Chemistry WebBook: \url{https://webbook.nist.gov/chemistry/}}$^]$, we choose an approximate value of $3.7 \times 10^8$ erg mol$^{-1}$ K$^{-1}$ for the molar heat capacity of the two gases \citep{chase1998}. On dividing this value respectively by the molar mass of molecular nitrogen and hydrogen molecules, the specific heat capacities are determined to be $\mathcal{C}_{N_2}\approx 1.3\times 10^7$ erg g$^{-1}$ K$^{-1}$ and $\mathcal{C}_{H_2}\approx1.83 \times 10^8$ erg g$^{-1}$ K$^{-1}$. If atmospheres dominated by CO$_2$ are evaluated instead, the specific heat capacity is lowered by a factor of $\sim 1.6$ compared to $\mathcal{C}_{N_2}$. As demonstrated later in Figure \ref{FigMPSpeed}, the molar mass does not play a significant role in our model for the inner $\sim 0.1$-$1$ kpc of the Milky Way.

By making use of this data, we have plotted $\Delta T$ as a function of $R$ in Figure \ref{FigHeatAtm} for an Earth-like planet. There are four curves in total because there exist two choices of the atmospheric composition and two values of $\epsilon$ in (\ref{modelheating}); the same reasoning is applicable to the other figures. It is apparent from inspecting Figure \ref{FigHeatAtm} that $\Delta T$ is considerable for distances of $R \lesssim 1$ kpc: this is partly due to the very long timescale over which the atmosphere is heated, which results in a substantial deposition of wind energy. As stated earlier, $\Delta T$ constitutes an upper bound since all the energy is modelled as being directed toward heating the atmosphere.

After the heating, the new temperature of the atmosphere is taken to be $T'$. If the atmosphere settles into a new state of thermal equilibrium, its molecules would follow the Maxwell-Boltzmann distribution. The most probable speed of the molecules ($v_\mathrm{mp}$) is easy to gauge in this instance \citep{EHK38}, and is evaluated as
\begin{equation}\label{vmp}
    v_\mathrm{mp} = \sqrt{\frac{2 k_B T'}{m_s}},
\end{equation}
where $m_s$ denotes the mass of a single molecule (either N$_2$ or H$_2$). The new temperature is defined to be $T' = T_0 + \Delta T$, where $T_0$ is the initial temperature; when $\Delta T$ is large, we see that $T' \approx \Delta T$. Although it is customary to select an isothermal temperature of $\sim 250$ K for the atmosphere \citep{DJJ99}, we adopt a slightly higher value of $T_0 \approx 273$ K; this modest discrepancy does not affect our analysis much, especially when $\Delta T \gg T_0$ is valid.

We have plotted $v_\mathrm{mp}$ as a function of the distance $R$ in Figure \ref{FigMPSpeed}. There are two notable features that stand out on scrutinising this figure. First, the velocity initially decreases with $R$, but then flattens out and reaches an ``equilibrium'' value (amounting to $\Delta T \rightarrow 0$). Second, and perhaps more pertinently, $v_\mathrm{mp}$ is higher than the escape velocity of the Earth (approximately equal to $11.2$ km/s) for distances of $\sim 0.1$ kpc in the energy-driven case, but this statement does not hold true for the momentum-driven winds. Hence, in the former situation, it is conceivable that Earth-like planets at distances of $\lesssim 0.1$ kpc are readily depleted of their atmospheres via thermal escape since a sizeable fraction of the molecules would possess velocities above that of the escape speed.

\subsection{Energy-limited atmospheric escape}

Next, we consider a different mechanism for atmospheric escape, which is modelled along the lines of the well-known paradigm of energy-limited hydrodynamic escape \citep{CK17,Ow19,ML21}. In this scenario, which is often applied to X-rays and extreme ultraviolet (XUV) radiation, the incident energy is transmuted into the kinetic energy of the constituent atmospheric particles, thereby enabling the latter to escape the planet. The impact of hydrodynamic escape mediated by AGN has already been investigated by \citet{BT17}, \citet{FL18}, and \citet{WKB}. Motivated by this approach, we will construct a similar energy-limited formulation with the wind energy replacing the electromagnetic energy. 

The total energy in the wind is given by $\dot{\mathcal{E}}_k \Delta t_\mathrm{Salp}$ and the fraction intercepted by the planet is $R_p^2/(4 R^2)$, as explained previously. To calculate the maximal amount of atmosphere depleted (denoted by ${M}_\mathrm{lost}$), we will suppose that all of the incident wind energy is converted into the kinetic energy of the escaping particles, which therefore require the escape velocity $v_e$. By assembling this information together, we end up with
\begin{equation}\label{EnBalance}
    \frac{R_p^2}{4 R^2} \dot{\mathcal{E}}_k \Delta t_\mathrm{Salp} \approx \frac{1}{2}{M}_\mathrm{lost}v_e^2.
\end{equation}
We will further rewrite the escape velocity as
\begin{equation}
    v_e \equiv \sqrt{\frac{2 G M_p}{R_p}} = 
    R_p \sqrt{\frac{8}{3}\pi G \rho_p},
\end{equation}
where $M_p$ and $\rho_p$ are the mass and mean density of the planet, respectively. On plugging this equation into (\ref{EnBalance}), we find that
\begin{equation}\label{massloss}
    M_\mathrm{lost}=\frac{3}{16 \pi G \rho_p}\frac{\epsilon L_\mathrm{Edd}\Delta t_\mathrm{Salp}}{R^2},
\end{equation}
implying that $M_\mathrm{lost}$ scales inversely with the square of the distance and that it only depends on the planet's mean density. We will utilise $\rho_p \approx \rho_\oplus \approx 5.5$ g/cm$^{3}$, i.e., we suppose that the planet's average density is equal to that of the Earth.

\begin{figure}
\includegraphics[width=8.5cm]{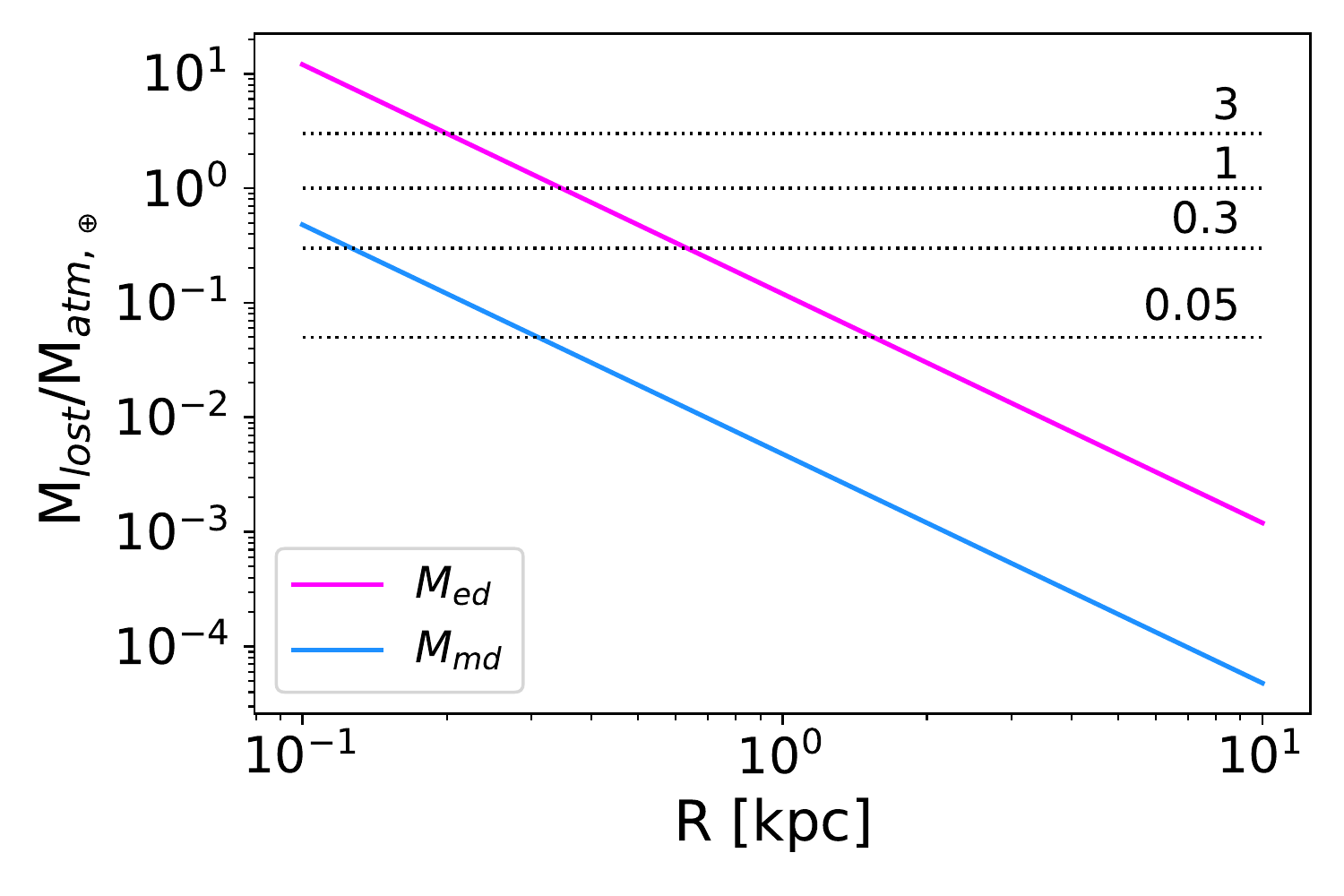} \\
\caption{The upper limit on the atmospheric mass lost (relative to Earth's atmospheric mass) because of energy-limited wind-mediated escape as a function of the distance $R$ from the Galactic centre (in kpc). The labels `ed' and `md' are indicative of the energy- and momentum-driven scenarios, respectively (see Section \ref{SecAGNPhen}). The horizontal lines denote the atmospheric mass expressed in terms of the mass of Earth's atmosphere.}
\label{FigMassLoss}
\end{figure}

The maximal atmospheric mass depleted versus the distance from the centre of the Galaxy is plotted in Figure \ref{FigMassLoss}. Since $\epsilon$ is higher for energy-driven winds than their momentum-driven counterparts, it is not surprising that $M_\mathrm{lost}$ is commensurately elevated. For the energy-driven scenario, we determine that a distance of $R \sim 0.4$ kpc might suffice to ensure that $M_\mathrm{lost}$ is nearly equal to the mass of Earth's atmosphere. Although this distance is small compared to the Sun's location, it should nevertheless be recognised that as many as $\gtrsim 10^9$ stars might reside in this region \citep{RRDP}, which is a large number in absolute terms. Energy-driven winds can stimulate the removal of Mars-like atmospheres at distances of $> 1$ kpc because the Martian atmosphere is merely $\sim 0.5$ percent as massive as that of Earth's atmosphere.

It is tempting to compare the atmospheric losses driven by AGN outflows versus stellar winds. This comparison is, however, not straightforward for two reasons. First, atmospheric escape powered by stellar winds can take place, in principle, over the lifetime of the star, whereas AGN outflows are operational over the relatively transient Salpeter timescale of $\sim 10$-$100$ Myr \citep{Shen13}. Second, the rates of atmospheric escape are not merely sensitive to planetary properties but also to stellar parameters such as mass and rotation rate. With these caveats in mind, we note that the timescale for depleting Earth's atmospheric mass is potentially $\sim 3.7 \times 10^8$ yr for Proxima b \citep{GGD17,DLMC} and orders of magnitude higher for Earth \citep{SEH01}, if the Sun were to hypothetically survive that long. In contrast, the Salpeter timescale is evidently faster, implying that temperate rocky planets at distances of $\lesssim 0.1$-$1$ kpc from the Milky Way centre may be conceivably dominated by AGN effects when it comes to non-thermal atmospheric escape mediated by winds.

Before moving on, we caution that energy-limited escape is only one of many processes that drive the depletion of atmospheres. We have not examined the plethora of non-thermal ion escape mechanisms (e.g., polar wind) known to function as one of the primary drivers of atmospheric loss on weakly magnetised planets in our Solar system and exoplanets \citep{BBM16,DLMC,DJL18,DLM18,LL19,ABC20}. An in-depth treatment of this crucial issue would necessitate performing sophisticated multi-species magnetohydrodynamics (MHD) simulations, which lies manifestly beyond the scope of our paper.

\subsection{Possible consequences of atmospheric depletion}\label{SSecAtDepCon}
The loss of an atmosphere is presumably inimical to surficial habitability in multiple respects. At the outset, we caution that even the total removal of an atmosphere does not translate to a complete elimination of life: certain refugia may still persist, especially in the deep subsurface \citep{WCW98,MLD18}, where chemosynthetic life could survive relatively unscathed. Yet, at the same time, it must be recognised that these deep biospheres are unlikely to generate readily detectable biosignatures.

Bearing the above provisos in mind, some of the most notable detriments are: (1) allowing substantial fluxes of UV and ionising radiation to penetrate to the surface, thereby instigating extensive biological damage (see also Section \ref{SSecDamage}), (2) suppressing the existence of liquid water bodies, as enforced by the phase diagram of water in the absence of surface pressure from an atmosphere, and (3) eliminating the prospects of detecting atmospheric biosignatures since they require, by definition, the existence of an atmosphere \citep{SKP18,FAD18,ML21}.

Before unpacking (1), (2), and (3) further, it is helpful to bear the key aspects of Figures \ref{FigMPSpeed} and \ref{FigMassLoss} in mind. From Figure \ref{FigMPSpeed}, it is apparent that the most probable speed of molecules -- due to atmospheric heating by the AGN winds -- exceeds that of the escape speed for Earth-sized worlds provided that $R \lesssim 0.1$ kpc for the energy-driven case. Therefore, terrestrial planets situated at such distances from the centre of the Milky Way are liable to losing a substantial fraction of their atmosphere via thermal escape mechanisms.

On the other hand, Figure \ref{FigMassLoss} illustrates the upper bound on the atmospheric mass lost via processes akin to energy-limited hydrodynamic escape, albeit involving energetic particles in lieu of XUV radiation. From this plot, we notice that the atmospheric mass depleted can equal that of Earth's atmosphere for energy-driven winds up to distances of $\sim 0.4$ kpc. Thus, Figures \ref{FigMPSpeed} and \ref{FigMassLoss} jointly indicate that Earth-sized planets with an atmosphere of $\sim 1$ bar are susceptible to being completely depleted up to distances of order $0.1$ kpc. We will now elaborate on the potential impediments to surface habitability anticipated from the loss of an atmosphere, which were adumbrated two paragraphs before.

First, in the absence of an atmosphere, there would be a massive increase in the UV-B and UV-C radiation reaching the surface. More precisely, for a planet analogous to modern Earth around a Sun-like star, the UV-B and UV-C top-of-atmosphere fluxes are $\Phi_{B} \approx 8.63 \times 10^3$ erg cm$^{-2}$ s$^{-1}$ and ${\Phi}_{C} \approx 3.38 \times 10^3$ erg cm$^{-2}$ s$^{-1}$ \citep[Tables 5 and 6]{RSKS}. In contrast, the UV-B and UV-C fluxes at the surface of modern Earth are, respectively, considerably lower at $8 \times 10^2$ erg cm$^{-2}$ s$^{-1}$ and $2.3 \times 10^{-13}$ erg cm$^{-2}$ s$^{-1}$ \citep[Tables 5 and 6]{RSKS}.

Likewise, when an Earth-like atmosphere (i.e., with a pressure of $\sim 1$ bar and column density of $\sim 10^3$ g cm$^{-2}$) is depleted, a substantial enhancement in the dose rate due to the higher flux of cosmic rays (and other energetic particles) penetrating to the surface is predicted. In particular, for an unmagnetized Earth-analog, numerical modelling suggests that reducing the atmospheric column density by a factor of $\sim 10$ boosts the biological dose rate by a factor of $\sim 850$ \citep[Table 2]{GTS16}.

There is, however, one crucial feature worth emphasising in connection with the elevated radiation and particle doses. The process of atmospheric escape would occur during the lifetime of the active phase of the SMBH, which corresponds to the Salpeter timescale defined in (\ref{SalpTime}). As this timescale is $\sim 10$-$100$ Myr \citep{Shen13}, it is apparent that the boost in electromagnetic radiation and particle fluxes is not abrupt. It is conceivable that organisms might be able to adapt to the thinning atmosphere by retreating to suitable refugia, such as subterranean and submarine environments, or evolve suitable adaptations \citep{CK99}.

Of greater concern is the second limitation highlighted previously, namely, the inability of water to exist as a liquid in the absence of finite pressure. Hence, if a rocky planet were to completely lose its atmosphere, the surface pressure would approach zero and thereupon rule out the presence of long-standing liquid water bodies. As liquid water is one of the requirements for life-as-we-know-it \citep{CBB16}, the absence of an atmosphere would, in turn, presumably rule out surface habitability, although subsurface habitats are feasible in principle, as remarked earlier.

Lastly, even if subsurface life were to survive on airless worlds, the absence of an atmosphere and surface-based life (which arises from the lack of liquid water) ostensibly removes the chief types of biosignatures detectable by telescopes -- to wit, atmospheric biosignatures (due to lack of an atmosphere) and surface biosignatures (because of the lack of surficial life). Hence, if an atmosphere is non-existent, the chances for identifying biomarkers on that world are strongly suppressed even from a theoretical perspective.

Although we have hitherto dwelt on the negative consequences of atmospheric loss, there is an important scenario where this process may lead to positive outcomes. As demonstrated in \citet{CFL18}, under the right circumstances, the photoevaporation of sub-Neptune-sized planets at $\sim 10$ pc due to XUV radiation emitted by Sagittarius A* can result in the formation of habitable rocky cores. In place of XUV radiation, it is plausible that energetic particles from UFOs could play a similar role. However, we do not address this matter herein since there are more uncertainties involved, especially concerning the characteristics of UFOs at such distances and their impact(s) on sub-Neptunes.

\section{Ozone depletion and consequences}\label{SecOzone}

\begin{figure}
\includegraphics[width=8.5cm]{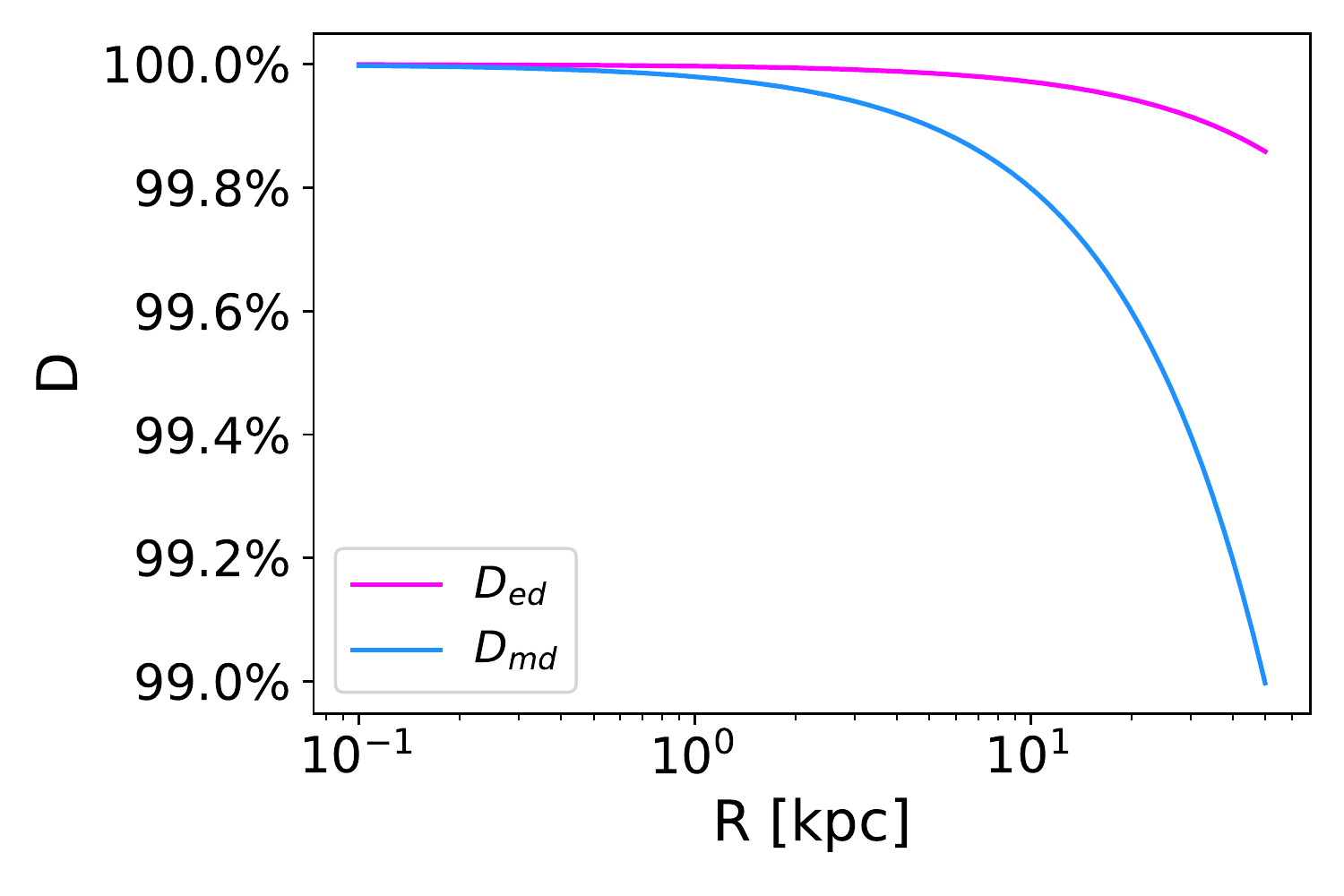} \\
\caption{The upper limit on the ozone depleted in an Earth-like atmosphere (denoted by $D$) because of NO$_x$ production stimulated by the energetic AGN wind as a function of the distance $R$ from the Galactic centre (in kpc). The labels `ed' and `md' refer to the energy- and momentum-driven scenarios, respectively (see Section \ref{SecAGNPhen}).}
\label{FigOzDep}
\end{figure}

One of the ubiquitous effects of high-energy particles -- whether they are generated by means of stellar flares and superflares, supernovae, gamma-ray bursts, or other phenomena -- is that they may contribute to ozone depletion of Earth-like atmospheres \citep{MT11,AM14,LL17}. If we consider stellar superflares on Sun-like stars, semi-analytical modelling by \citet{LL17} suggests that sufficiently large events that can completely erode the ozone layers of Earth-like planets might transpire only once every $\mathcal{O}(10)$ Myr. On the other hand, as per numerical simulations, active M-dwarfs that produce regular flares of $\sim 10^{32}$ erg once every month may cumulatively instigate the loss of an ozone layer in $\sim 10^7$ yr on Earth-like planets in their habitable zones \citep[Table 2]{TSM19}.

We caution that the great diversity in stellar properties (e.g., activity and mass) makes it very difficult to designate a ``typical'' timescale for ozone depletion arising from stellar flares, which are themselves varied in scope. This issue is complicated further by the complex time-dependent response of planetary atmospheres to the electromagnetic radiation and energetic protons (and electrons) associated with flare activity, as revealed by the state-of-the-art simulations performed by \citet{CZY21} to assess the effects of atmospheric chemistry triggered by flares.

As intimated previously, UFOs have speeds of $\sim 0.1 c$ and even after interacting with the ISM they exhibit speeds of $\mathcal{O}(1000)$ km/s \citep{MAB09,TCR11,TMV15,VPB18}. Hence, it is instructive to explore whether outflows from AGN with high enough speeds (e.g., akin to UFOs) can likewise contribute to ozone depletion. In principle, high-energy particles may also lead to some positive outcomes such as the synthesis of prebiotic molecules, but we do not evaluate the latter because this process entails more unknowns and uncertainties \citep{LDF18}. In contrast to the previous section (Section \ref{SecAtmos}), we focus exclusively on modern Earth-like atmospheres composed of N$_2$ and O$_2$ herein (and not hydrogen-dominated atmospheres) since we wish to investigate the prospects for ozone depletion by nitrogen oxides \citep{PJC71}, which could be formed during the interactions of high-velocity AGN outflows with Earth-like atmospheres as described hereafter.

The ozone layer on Earth is believed to have become prominent only when the Great Oxygenation Event (GOE) occurred $\sim 2.4$ Ga \citep{JK87,LDP21}. As our stated objective is to consider planets similar to modern Earth, we do not consider analogues of Hadean-Archean Earth, which was predominantly anoxic and harboured higher concentrations of CO$_2$ compared to today. In analysing ozone depletion, we focus on pathways involving nitrogen oxide species NO$_x$ and not hydrogen oxides HO$_x$ (or chlorine oxides ClO$_x$). The reason is that the influx of high-energy particles inherent in fast AGN outflows might stimulate the production of NO$_x$ as described below. In contrast to NO$_x$, whose synthesis is enabled by high-energy particles \citep{SS99}, the major sources for HO$_x$ are not based on this pathway (see \citealt{WM74}). Since chemical species other than NO$_x$ that facilitate ozone depletion are neglected in this paper, we note that the calculations may constitute a relatively conservative assessment of this phenomenon.

It is well-known that energetic particles, like those produced (in)directly during solar flares, promote the formation of NO$_x$ \citep{PJC71,CIR75}, which can destroy O$_3$ through the following catalytic reactions:
\begin{equation}\label{OzDes}
    \mathrm{NO} + \mathrm{O_3} \xrightarrow[]{} \mathrm{NO_2} + \mathrm{O_2},
\end{equation}
\begin{equation}
    \mathrm{NO_2} + \mathrm{O} \xrightarrow[]{} \mathrm{NO} + \mathrm{O_2}. 
\end{equation}
To elaborate further, the abundance of NO$_x$ (NO and NO$_2$) may increase by virtue of an enhanced flux of high-energy particles that are loosely analogous to cosmic rays and solar energetic particles. The latter duo are documented to produce ion pairs in the atmosphere, which thereupon react with molecules such as N$_2$ and dissociate them \citep{CIR75,Dart11}. The ensuing free nitrogen atoms are responsible for nitric oxide (NO) creation and destruction through the reactions:
\begin{equation}\label{Create}
    \mathrm{N} + \mathrm{O_2} \xrightarrow[]{}\mathrm{NO} + \mathrm{O},
\end{equation}
\begin{equation}\label{Destroy}
    \mathrm{N} + \mathrm{NO} \xrightarrow[]{} \mathrm{N_2} + \mathrm{O}.
\end{equation}
and the net NO so formed in this fashion could catalyse the depletion of ozone along the lines implied in (\ref{OzDes}).

A full-fledged treatment of the production of NO$_x$ and their consequent destruction of ozone would require full-fledged numerical simulations incorporating atmospheric chemistry. We will, instead, adopt the semi-analytical framework presented in \citet{ES95} (see also \citealt{MAR74}), which has proven to be fairly accurate in modelling ozone depletion from supernovae \citep{BW17}, when compared against more comprehensive treatments of this subject; in other words, the formalism is sufficient for heuristic purposes. Additional details pertaining to the subsequent discussion can be found in \citet{ES95}. 

\subsection{Energetic particles and nitrogen oxide production}
The production rate of NO ($R_\mathrm{NO}$) associated with an increase in the flux of energetic particles is given by
\begin{equation}\label{RNO}
    R_\mathrm{NO}=R_0 \frac{\Phi}{\Phi_0}\frac{10+y_0}{10+y} \mskip\thinmuskip \text{molecules cm}^{-2}\text{yr}^{-1},
\end{equation}
where $R_0 \approx 9 \times 10^{14}$ molecules cm$^{-2}$ yr$^{-1}$ signifies the rate at which the ambient flux of cosmic rays generates NO, suitably averaged over time and latitude and integrated over altitude; $\Phi$ represents the averaged energy flux attributable to the AGN outflow particles; and  $\Phi_0\approx 9 \times 10^4$ erg cm$^{-2}$ yr$^{-1}$ encapsulates the energy flux at the Earth's surface from cosmic rays. The last term on the right-hand side is an efficiency ratio that accounts for the creation and destruction of NO through the reactions (\ref{Create}) and (\ref{Destroy}).

As previously remarked, ion pairs instigate the production of NO$_x$. The net quantity of NO generated per ion pair is found to be $10/(10+y)$ \citep{MAR74}, where $y$ represents the abundance of stratospheric NO measured in units of parts per billion (ppb). Here, in the same vein as \citet{ES95}, we have assumed that the synthesis of NO dominates over that of NO$_2$. Therefore, the final term in (\ref{RNO}) is the ratio of the net NO production during the outflow-stratosphere interaction, namely $10/(10+y)$, and the net NO generated by ambient cosmic rays, namely $10/(10+y_0)$, where the variable $y_0$ embodies the background NO abundance arising from cosmic rays, which we take to be $y_0=3$ ppb. In (\ref{RNO}), the unknown parameter is $y$, which is investigated below. 

The NO concentration whose production is stimulated by the AGN outflow is modelled as
\begin{equation}\label{ydef}
    y=\frac{R_\mathrm{NO} \tau }{\sigma_\mathrm{strat}}\times 10^9,
\end{equation}
where $\sigma_\mathrm{strat}$ represents the stratospheric column density, chosen to be $\sigma_\mathrm{strat}= 5 \times 10^{23}$ molecules cm$^{-2}$ based on the Earth. The factor of $10^9$ on the right-hand side appears because $y$ is expressed in units of ppb. The timescale $\tau$ was interpreted in \citet{ES95} as the residence time for NO in the stratosphere, and a value of $4$ yr was specified; note that the lifetime of NO$_x$ in the troposphere is much shorter, viz., on the order of days \citep{PJC79}.

Unlike phenomena such as superflares, gamma-ray bursts, and supernovae, AGN do not engender a high fluence of energetic particles in a relatively short period. To put it differently, the perturbation applied to the stratosphere is not transient -- in fact, the ``forcing'' applied to the system continues over the lifetime of the AGN, the latter of which is approximated by the Salpeter timescale. Therefore, we can calculate the maximal ozone depletion feasible if we posit that $\tau$ is set equal to $\Delta t_\mathrm{Salp}$. In doing so, the assumption is that NO is being continuously produced in the stratosphere by AGN-driven winds and that the losses are not as prominent.

On substituting (\ref{ydef}) into (\ref{RNO}) and inverting the latter equation to solve for $y$, we end up with
\begin{equation}\label{QuadEy}
    y^2+10y-R_0\frac{\Phi}{\Phi_0}\cdot\frac{(10+y_0)\Delta t_\mathrm{Salp} \cdot 10^9}{\sigma_\mathrm{strat}}=0.
\end{equation}
Once $y$ has been estimated by solving this equation, the NO abundance can be employed to determine the expected degree of ozone loss. The formalism resembles that of \citet{MAR74} and \citet{ES95} in computing the ratio $F$ of stratospheric ozone abundance in the perturbed ($[\mathrm{O}_3]$) and unperturbed ($[\mathrm{O}_3]_0$) cases; the ambient cosmic rays are responsible for the latter, whereas the former is driven by the AGN outflows. As per the aforementioned framework, this ratio is derived from
\begin{equation}\label{Ozdep}
     F=\frac{[\mathrm{O}_3]}{[\mathrm{O}_3]_0}=\frac{\sqrt{16+9X^2}-3X}{2},
\end{equation}
where the dimensionless variables $X$ quantifies the ratio of the perturbed and unperturbed $NO$ abundances as follows:
\begin{equation}\label{xno}
    X=\frac{[\mathrm{NO}]}{[\mathrm{NO}]_0}=\frac{y_0 + y}{y_0}=\frac{3+y}{3},
\end{equation}
where the last equality is obtained by invoking the fiducial value of $y_0=3$ ppb introduced before. Once the value of $F$ is known, the fractional ozone depletion $D \equiv 1-F$ can be inferred accordingly.

The missing ingredient in (\ref{QuadEy}), and thence in (\ref{Ozdep}) and (\ref{xno}), is the energy flux associated with the AGN wind: $\Phi$ is found by dividing the kinetic power by the spherical area of $4\pi R^2$, and the subsequent result is lowered by an extra factor of $4$. The latter stems from the fact that the flux is intercepted by a cross-sectional area of $\pi R_p^2$, but must be distributed over the planet's total surface area of $4\pi R_p^2$. With these simplifications, we arrive at the average flux of
\begin{equation}\label{normalflux}
    \Phi =\frac{\dot{\mathcal{E}}_k}{16\pi R^2},
\end{equation}
and this energy flux must be converted into the units of erg cm$^{-2}$ yr$^{-1}$ to preserve the same units as $\Phi_0$.

The outcome of the ozone depletion modelling, under the given set of assumptions, is depicted in Figure \ref{FigOzDep}. On inspecting the figure, it is evident that $D > 99$ percent even up to the outer boundary of the Milky Way. It would appear, \emph{prima facie}, that AGN outflows can efficiently destroy virtually all of the stratospheric ozone on Earth-like planets in the Galaxy. We reiterate, however, that the ozone depletion $D$ is an upper bound, and a loose one at that, since NO production was evaluated over the entire Salpeter timescale. To undertake a more realistic treatment, we will adopt a different strategy described below.

\begin{figure}
\includegraphics[width=8.5cm]{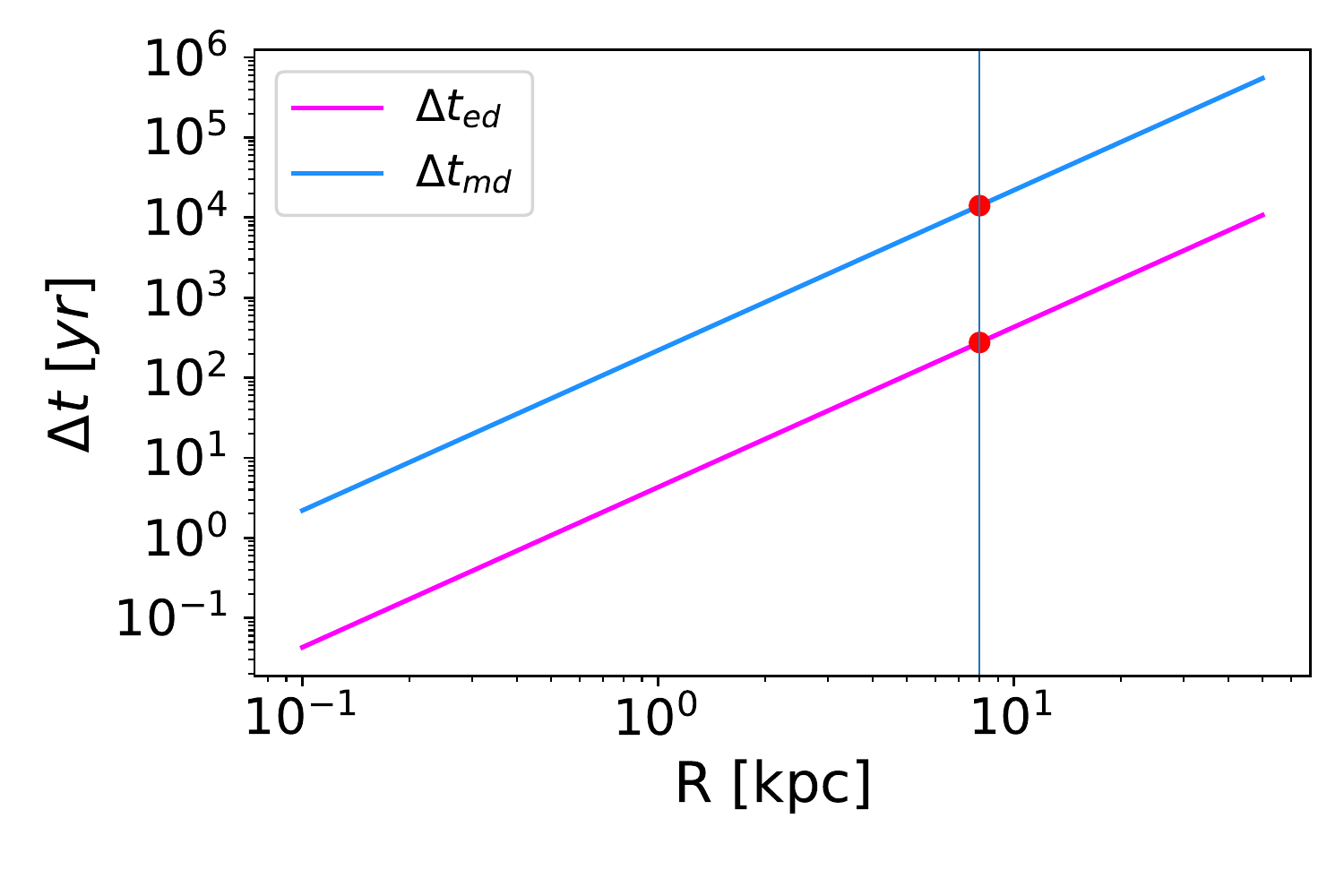} \\
\caption{The timescale over which NO must be continuously active in the atmosphere of an Earth-like planet to effectuate $90$ percent ozone depletion as a function of the distance $R$ from the Galactic centre (in kpc). The labels `ed' and `md' are representative of the energy- and momentum-driven scenarios, respectively (see Section \ref{SecAGNPhen}).}
\label{Fig90PerOz}
\end{figure}

By applying the same formalism, it is feasible to determine the time required to instantiate a depletion of $90$ percent of atmospheric ozone. Even an ozone depletion of $\sim 30$ percent could lead to a doubling of the UV-B flux at the surface and might trigger a mass extinction \citep{Gehrels2003,TMJ05,MT11}. In contrast, an ozone depletion of $90$ percent would increase the UV-B flux by several times and thus may cause substantial damage to ecosystems. By deploying $D \equiv 1 - F = 90$ percent in conjunction with (\ref{normalflux}), (\ref{QuadEy}), (\ref{xno}), and (\ref{Ozdep}), the timescales for $90$ percent ozone depletion are approximated by
\begin{equation}\label{Delted}
    \Delta t_\mathrm{ed} \approx 4.5\,\mathrm{yr}\,\left(\frac{R}{1\,\mathrm{kpc}}\right)^2
\end{equation}
\begin{equation}\label{Deltmd}
    \Delta t_\mathrm{md} \approx 220\,\mathrm{yr}\,\left(\frac{R}{1\,\mathrm{kpc}}\right)^2
\end{equation}
where the subscripts `ed' and `md' refer to energy- and momentum-driven outflows, respectively. In other words, if the stratospheric NO generated from AGN outflows has an effective residence equal to the expressions presented above, one may expect around $90$ percent of the stratospheric ozone to be eliminated. If we set $\Delta t_\mathrm{ed} \approx 4$ yr as indicated earlier, we see that ozone depletion could be a serious issue even up to distances of $\sim 1$ kpc.

The results from the exact numerical calculations of $\Delta t_\mathrm{ed}$ and $\Delta t_\mathrm{md}$ are illustrated in Figure \ref{Fig90PerOz}. Along expected lines, we notice that distances further away from the Galactic centre would require longer timescales over which NO is functional in the stratosphere. If we select $R \approx 8$ kpc, which signifies the distance of the Earth from the centre of the Milky Way, the time necessary for $90$ percent ozone depletion as per our modelling is $\Delta t_\mathrm{ed} \approx 3 \times 10^2$ yr for a UFO or an energy-driven outflow and it increases to $\Delta t_\mathrm{md} \approx 1.4 \times 10^4$ yr if the outflow were momentum-driven instead.

Let us focus on the energy-driven outflow as it requires a relatively lower residence time for NO. In contrast to the timescale of $\sim 4$ yr chosen for Earth \citep{ES95}, the preceding paragraph indicates that a much longer lifetime of $\sim 300$ yr is necessary. While this value does not seem plausible for Earth-like worlds around Sun-like stars, it might be feasible for such planets orbiting quiescent late-type M-dwarfs on account of the comparative paucity of UV radiation, although the specifics must be evaluated on a case-by-case basis. For methane, \citet{SKM05} demonstrated that the photochemical lifetime is enhanced by a factor of $\sim 17$ on M-dwarf Earth-analogues with respect to the Earth, and similar results were obtained for other gases such as N$_2$O \citep{GGG13}; see \citet{JLG17} for a succinct review.

\subsection{Possible impact on biomolecules and life}\label{SSecDamage}
On the basis of the prior analysis, it is clear that some fraction of planets in the Milky Way would experience major ozone depletion. In particular, our salient findings are encapsulated in Figures \ref{FigOzDep} and \ref{Fig90PerOz}. If the existence of nitrogen oxides produced by UFOs is sustained throughout the lifetime of the AGN, then all Earth-like planets in the Milky Way would become virtually devoid of their ozone layers (see Figure \ref{FigOzDep}). In contrast, as per Figure \ref{Fig90PerOz}, if the lifetime of NO$_x$ is merely a few years (i.e., akin to Earth), worlds up to distances of $\lesssim 1$ kpc are susceptible to experiencing ozone depletion of $\sim 90$ percent in the energy-driven paradigm.

In light of our preceding exposition, we will now assess the biological damage wrought by this process. One of the immediate consequences is that much of the UV radiation from the host star at wavelengths $> 200$ nm will reach the surface, which can engender a variety of negative effects such as damage to biomolecules (e.g., DNA), inhibition of nitrogen fixation, and suppression of photosynthesis \citep{CK99,Dart11,ML21}. We will focus our attention on Earth-like planets and Sun-like stars; it is straightforward to generalise the same approach to K- and M-dwarfs, among others.

The TOA UV-B and UV-C fluxes at Earth contributed by the modern Sun are respectively $\Phi_{B}$ and ${\Phi}_{C}$, delineated in Section \ref{SSecAtDepCon}. As ozone is responsible for the absorption of $\sim 90$ percent of UV-B radiation, in its near-complete absence, it is reasonable to surmise that the majority of $\Phi_B$ will penetrate to the surface; hence, we adopt this value for UV-B radiation. On the other hand, molecules aside from ozone absorb UV-C radiation, owing to which the TOA flux would not be the same as the surficial UV-C flux. In the absence of an ozone layer, we approximate the UV-C flux at the surface with that estimated for Archean Earth (which lacked ozone) and specify $\Phi'_{C}=871$ erg cm$^{-2}$ s$^{-1}$ \citep[Table 6]{RSKS}.

The UV-C fluence required to kill $90$ percent (the D$90$ dose) of the radioresistant extremophile \emph{Deinococcus radiodurans} is $\mathcal{F}_C \approx 5.5 \times 10^5$ erg cm$^{-2}$ \citep{GOP95}. By employing the UV-C flux $\Phi'_C$ from the preceding paragraph, the characteristic extinction timescale $\Delta t_\mathrm{kill}$ is estimated to be
\begin{equation}\label{KillTimeC}
    \Delta t_\mathrm{kill} \approx \frac{\mathcal{F}_C}{\Phi'_{C}} \approx 6.3 \times 10^2\,\mathrm{s}.
\end{equation} 
In contrast, if we presume that the UV-C surficial flux is highly ameliorated somehow by certain molecular species in the atmosphere, the UV-B radiation still represents a potent threat for organisms. As before, we consider the extremophile \emph{D. radiodurans} and analyse its capacity to withstand UV-B radiation. The UV-B fluence necessary to kill \emph{D. radiodurans} was investigated by \citet{KGK21}. This study concluded that complete killing was achievable when this species was subjected to a UV-B flux of $2 \times 10^3$ erg cm$^2$ s$^{-1}$ -- which is comparable to the value of $\Phi_{B}$ -- for a period of $16$ h, thereby implying a lethal UV-B dose of $\mathcal{F}_B \approx 1.2 \times 10^8$ erg cm$^2$. If a large fraction of TOA UV-B flux reaches the surface, the extinction timescale is roughly given by
\begin{equation}\label{KillTimeB}
    \Delta t_\mathrm{kill} \approx \frac{\mathcal{F}_B}{\Phi_{B}} \approx 1.3 \times 10^4\,\mathrm{s}.
\end{equation}
Hence, as per (\ref{KillTimeC}) and (\ref{KillTimeB}), it would seem as though even radioresistant organisms like \emph{D. radiodurans} could be rendered extinct in short timescales (minutes to hours) as a consequence of the high ozone depletion caused on some planets by AGN winds. 

However, it is vital to appreciate that high UV radiation does not, by itself, spell doom for all lifeforms on a particular world. Even \emph{sans} an ozone shield, it is well-known that microbial ecosystems were thriving in the Archean eon \citep{Knoll,ML21}. Moreover, shielding accorded by water and soil, \emph{inter alia}, can protect organisms in high-UV environments. Last, but not least, organisms have evolved a diverse array of screening compounds to mitigate the effects of UV radiation \citep{CK99}. All of these facets might mitigate the damage wrought by the elevated UV fluxes at the surface in the sudden absence of ozone.

Even in the worst-case scenario where widespread extinction of species (possibly a mass extinction) is triggered by the rapid loss of ozone, it is still conceivable that the biosphere may recover over time provided that there are no further disruptions that hamper this course of events. The Permian–Triassic (P-Tr) extinction event -- which constitutes the largest mass extinction in the Phanerozoic eon, and was potentially characterised by the extinction of $\sim 80$ percent of marine species \citep{SMS16} -- was subsequently followed a gradual recovery that required $\lesssim 10$ Myr \citep{SB08,CB12}. A prediction of the recovery timescale is impractical because it depends on both the AGN properties as well as the nature of the putative ecosystems, which are unknown. However, if we go by the example of the P–Tr extinction, it does not seem altogether impossible for the biosphere to reattain stability and complexity after a suitable timescale.

\section{Discussion and Conclusions}\label{SecConc}

\begin{table*}
\begin{minipage}{155mm}
\caption{Potential maximum Galactic distances up to which effects of ultra-fast outflows from AGN are significant}
\label{EffAGN}
\vspace{0.1 in}
\begin{tabular}{|c|c|c|}
\hline 
Effect & Momentum-driven case (in kpc) & Energy-driven case (in kpc)\tabularnewline
\hline 
\hline 
Atmospheric escape arising from thermal heating & N/A & $\sim 0.2$ kpc \tabularnewline
\hline 
Energy-limited hydrodynamic-like atmospheric escape & $< 0.1$ kpc & $\sim 0.4$ kpc\tabularnewline
\hline 
Major ozone depletion due to nitrogen oxide formation & $\sim 0.15$ kpc & $\sim 1$ kpc\tabularnewline
\hline 
\end{tabular}
\medskip

{\bf Notes:} The distances are measured with respect to the centre of the Milky Way, which harbours the SMBH Sagittarius A*. The momentum- and energy-driven cases are described in Sections \ref{SSecMomDr} and \ref{SSecEnDr}, and correspond to winds wherein the conservation of momentum and energy are respectively valid. The term ``N/A'' is introduced when the effects in question are substantial only up to relatively negligible distances.
\end{minipage}
\end{table*}

Although the relevance of SMBH activity in regulating habitability on galactic scales has garnered attention, modern studies have not rigorously investigated the role of AGN winds and outflows in mediating galactic habitability. Hence, in this work, we sought to rectify this important lacuna using the Milky Way as a proxy.

After a brief introduction to the appropriate phenomenology of AGN outflows in Section \ref{SecAGNPhen}, we embarked on an exploration of the deleterious ramifications stemming from this phenomenon. In Section \ref{SecAtmos}, we scrutinised the extent of atmospheric heating and the accompanying thermal escape. Our analysis suggests that the latter becomes prominent at distances on the order of $0.1$ kpc for energy-driven winds, in which the energy of the outflow is transferred to the post-shock winds. By drawing on a model analogous to energy-limited hydrodynamic escape, we showed that energy-driven winds can facilitate cumulative atmospheric losses amounting to that of Earth's atmosphere at distances of $\lesssim 0.4$ kpc.

At first glimpse, it would appear as though the atmospheric escape powered by outflows is significant only at distances of $\mathcal{O}(0.1)$ kpc, whereas XUV radiation drives escape up to distances of $\mathcal{O}(1)$ kpc \citep{BT17}. However, it is crucial to recognise that the major mechanisms underpinning the escape of heavier species (e.g., nitrogen and oxygen) on terrestrial planets in our Solar system involve interactions with the solar wind and not just electromagnetic radiation \citep{BBM16}; the same is anticipated to hold true for exoplanets \citep{DLMC,DJL18}. Hence, when the standard model of energy-limited hydrodynamic escape due to XUV photons is not applicable, the energetic particles in AGN winds may serve as the primary instigators of atmospheric escape. Mars-like atmospheres, as seen from Figure \ref{FigMassLoss}, might be entirely depleted by energy-driven AGN winds even at large distances of $\gtrsim 1$ kpc.

In the next section (Section \ref{SecOzone}), we evaluated the impact of AGN winds on triggering ozone depletion in Earth-like atmospheres via production of nitrogen oxides, akin to how other high-energy phenomena (e.g., supernovae and stellar flares) can do the same. We showed that the upper bound on the ozone depletion is close to $100$ percent for the entirety of the Milky Way, but only under the limiting postulates that the synthesis of nitrogen oxides (which catalyse ozone depletion) occurs continuously over the lifetime of the AGN (viz., the Salpeter timescale), and that these compounds remain active in the atmosphere throughout the period.

By relaxing this assumption, we estimated the timescale over which nitrogen oxides must be functional in the atmosphere so as to effectuate $90$ percent ozone depletion. The ensuing results are presented in Figure \ref{Fig90PerOz} as well as (\ref{Delted}) and (\ref{Deltmd}). At distances of $\lesssim 1$ kpc, our analysis indicates that energy-driven AGN outflows may necessitate a timescale of a few years, which is potentially comparable to the lifetime of stratospheric nitric oxide in some Earth-like atmospheres. Thus, the impact of ozone depletion could prove to be substantial at distances of $\lesssim 1$ kpc. 

In the event that near-complete ozone depletion does transpire on some planets, the UV-B and perhaps the UV-C flux at the surface can become enhanced by nearly an order of magnitude. The abrupt elevation of UV radiation is predicted to cause widespread biological damage and might even give rise to a mass extinction. In Section \ref{SecOzone}, we demonstrated that even radioresistant extremophiles such as \emph{Deinococcus radiodurans} may be subjected to extinction over short timescales of minutes to hours. However, these statements are not equivalent to the extinction of life altogether, as there are multifarious environments that are mostly unaffected by the amplified UV radiation in the sudden absence of an ozone layer.

To sum up, in the case of the Milky Way, we determined that impacts on habitability arising from interactions of AGN outflows with planetary atmospheres are possibly significant at distances of $\lesssim 1$ kpc, as summarised in Table \ref{EffAGN}. We reiterate that we chose to employ the SMBH at the centre of the Milky Way as our proxy, but other galaxies have much larger SMBHs, as much as three orders of magnitude higher in mass than Sagittarius A* \citep[e.g.,][]{MMM12,KH13}. Many of the expressions derived herein are explicitly dependent on $\Delta t_\mathrm{Salp} L_\mathrm{Edd}$, and the latter is itself linearly proportional to $M_\mathrm{BH}$, as seen from (\ref{TotalEn}). Hence, the effects of AGN winds in quasar host galaxies are conceivably rendered increasingly predominant, owing to which they may consequently push the limits of the habitable zone to the galaxy outskirts. 

It is instructive to compare our results with prior analyses of the astrobiological potential of the Milky Way. In a seminal publication, \citet{LFG04} (see also \citealt{GBW01}) took factors such as metallicity and the frequency of supernovae into account and concluded that only the annular region of $7$-$9$ kpc would be well-suited for complex life, which was dubbed the Galactic Habitable Zone (GHZ). This finding has received support from some subsequent studies \citep[e.g.,][]{SGM17}, but is contradicted to varying degrees by others \citep{NP08,VSM16,FDC17}; for instance, simulations by \citet{FDC17} have yielded a much wider annulus of $2$-$13$ kpc.

Our current work in tandem with publications on the negative impacts of high-energy radiation from AGN \citep{BT17,LGB}, as well as the contributions from tidal disruption events \citep{PBL20}, collectively suggests that the innermost $\sim 1$ kpc of the Milky Way is not readily conducive to the existence of biospheres on the surfaces of rocky planets. This tentative conclusion should be seen as \emph{complementary} in scope to the papers cited in the prior paragraph because the latter did not incorporate SMBH activity in their modelling. We caution, however, that our analysis does not rule out the prospects for surficial life altogether owing to the uncertainties and caveats involved, some of which were encountered previously and others are delineated hereafter. Furthermore, the AGN phase is primarily operational over the Salpeter timescale, due to which life might be able to emerge and diversify on the surface -- perhaps by migrating from the subsurface or from other worlds \citep{CFL18,GLL18,LGB22} -- once this phase has ceased.

It is worth recalling that some of our calculations entailed the derivation of upper limits and/or the utilisation of simple prescriptions. Future research should endeavour to address these aspects, for example, by drawing on sophisticated atmospheric chemistry \citep{CZY21} and/or multi-species MHD \citep{DJL20} modelling to self-consistently account for ozone depletion and non-thermal atmospheric escape, respectively. On a related note, we have not tackled the question of how strong planetary magnetic fields might modulate the effects caused by ionised particles in AGN outflows. Current research appears to imply that planetary atmospheres are potentially more crucial than planetary magnetospheres (instantiated by magnetic fields) in regulating the near-surface fluxes of charged particles and their repercussions for habitability such as biological damage and atmospheric escape \citep{GTS16,DA17,DA20,DLML18,ML19}.

Likewise, we focused our attention on radiation driven UFOs (and their post-shock derivatives) because of their high speeds and prevalence. However, in light of the panoply of AGN winds and outflows, to say nothing of relativistic jets, our work should be viewed as a stepping stone for subsequent treatments along similar lines for various classes of AGN and their outflows. Moreover, the consideration of more sophisticated wind acceleration mechanisms (such as MHD-driven) could duly boost the energetics of the AGN outflows \citep[e.g.,][]{fuku2010,fuku2015}, thereby impacting a larger region of the galaxy. Lastly, our conclusions are applicable strictly to surficial habitability, and not to subsurface ecosystems that might actually be the most abundant in the Universe by as much as a few orders of magnitude \citep{Ling19,LL20}.

Lastly, a word regarding the empirical assessment of our findings is necessary. Since the maximum distance up to which the deleterious effects of AGN outflows are substantial is $\sim 1$ kpc as per our modelling, this would mean that the nearest such exoplanets are several kpc away from Earth. Hence, while the detection of these planets is feasible via microlensing \citep{GL92,VB18}, characterising them seems unlikely in the near-future. On longer timescales, the deployment of large telescopes at the Solar Gravitational Lens (SGL) \citep{TST19}, allied to promising advances in agnostic biosignatures \citep[e.g.,][]{BLG22}, might enable us to probe the innermost kiloparsec of the Milky Way and thereby gauge the impacts of past SMBH activity.

\section*{Data Availability Statement}
No new data were generated or analysed in support of this research.

\section*{Acknowledgements}
A.B.\ acknowledges support by the Italian Space Agency (ASI, DC-VUM-2017-034, grant number 2019-3 U.O Life in Space) and by grant number FQXi-MGA-1801 and FQXi-MGB-1924 from the Foundational Questions Institute and Fetzer Franklin Fund, a donor advised fund of Silicon Valley Community Foundation. The authors are grateful to the reviewer for the meticulous and insightful report, which helped us substantively improve the manuscript.

\bibliographystyle{mnras}
\bibliography{AGNHab}

\bsp	
\label{lastpage}
\end{document}